\documentclass[twocolumn,floatfix,aps,superscriptaddress]{revtex4}
\usepackage{graphicx}
\usepackage{amsmath}
\usepackage{amssymb}
\usepackage{bm}

\begin{document}

\title{Twisted Fermi surface of a thin-film Weyl semimetal}
\author{N. Bovenzi}
\affiliation{Instituut-Lorentz, Universiteit Leiden, P.O. Box 9506, 2300 RA Leiden, The Netherlands}
\author{M. Breitkreiz}
\affiliation{Instituut-Lorentz, Universiteit Leiden, P.O. Box 9506, 2300 RA Leiden, The Netherlands}
\author{T. E. O'Brien}
\affiliation{Instituut-Lorentz, Universiteit Leiden, P.O. Box 9506, 2300 RA Leiden, The Netherlands}
\author{J. Tworzyd{\l}o}
\affiliation{Institute of Theoretical Physics, Faculty of Physics, University of Warsaw, ul.\ Pasteura 5, 02--093 Warszawa, Poland}
\author{C. W. J. Beenakker}
\affiliation{Instituut-Lorentz, Universiteit Leiden, P.O. Box 9506, 2300 RA Leiden, The Netherlands}

\date{September 2017}
\begin{abstract}
The Fermi surface of a conventional two-dimensional electron gas is equivalent to a circle, up to smooth deformations that preserve the orientation of the equi-energy contour. Here we show that a Weyl semimetal confined to a thin film with an in-plane magnetization and broken spatial inversion symmetry can have a topologically distinct Fermi surface that is twisted into a \mbox{figure-8} --- opposite orientations are coupled at a crossing which is protected up to an exponentially small gap. The twisted spectral response to a perpendicular magnetic field $B$ is distinct from that of a deformed Fermi circle, because the two lobes of a \mbox{figure-8} cyclotron orbit give opposite contributions to the Aharonov-Bohm phase. The magnetic edge channels come in two counterpropagating types, a wide channel of width $\beta l_m^2\propto 1/B$ and a narrow channel of width $l_m\propto  1/\sqrt B$ (with $l_m=\sqrt{\hbar/eB}$ the magnetic length and $\beta$ the momentum separation of the Weyl points). Only one of the two is transmitted into a metallic contact, providing unique magnetotransport signatures.
\end{abstract}
\maketitle

\section{Introduction}
\label{intro}

The Fermi surface of degenerate electrons separates filled states inside from empty states outside, thereby governing the electronic transport properties near equilibrium. In a two-dimensional electron gas (2DEG) the Fermi surface is a closed equi-energy contour in the momentum plane. It is a circle for free electrons, with deformations from the lattice potential such as the trigonal warping of graphene or the hexagonal warping on the surface of a topological insulator \cite{warping}. These are all smooth deformations which do not change the orientation of the Fermi surface: The turning number is 1, meaning that the tangent vector makes one full rotation as we pass along the equi-energy contour. 

\begin{figure}[tb]
\centerline{\includegraphics[width=0.8\linewidth]{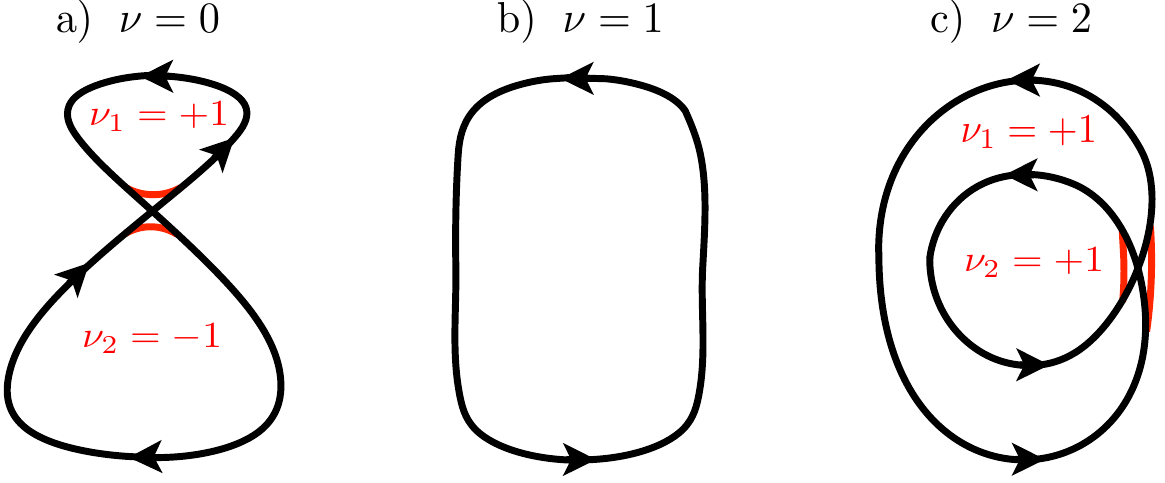}}
\caption{Three oriented contours (black curves) with turning number $\nu=0,1,2$. The red segments show the uncrossing deformation that removes a self-intersection without changing the total turning number $\nu=\sum_i\nu_i$.
}
\label{fig_turning}
\end{figure}

The turning number
\begin{equation}
\nu=\frac{1}{2\pi}\oint_\Gamma {\cal C}\,dl,\label{nudef}
\end{equation}
defined as the contour integral of the curvature ${\cal C}$ in units of $2\pi$, identifies topologically distinct deformations of the circle in the plane, socalled ``regular homotopy classes'' \cite{homotopy}. A theorem going back to Gauss \cite{Gauss} says that a contour $\Gamma$ with turning number $\nu$ has $s\geq\biglb||\nu|-1\bigrb|$ self-intersections and that the sum $|\nu|+s$ must be an odd integer. Fig.\ \ref{fig_turning} shows examples of contours with $\{\nu,s\}=\{0,1\}$, $\{1,0\}$, and $\{2,1\}$.

The turning number is preserved by any smooth deformation of the contour. This includes socalled ``uncrossing'' deformations \cite{homotopy}: As illustrated in Fig.\ \ref{fig_turning}, uncrossing breaks up a self-intersecting contour $\Gamma$ into a collection of nearly touching oriented contours $\Gamma_i$, with turning numbers $\nu_i$. The total turning number $\nu = \sum_i \nu_i $ is invariant against uncrossing deformations, which is another result due to Gauss \cite{Gauss}.

All familiar 2D electron gases belong to the $|\nu|=1$ universality class. Here we show that a thin-film Weyl semimetal with an in-plane magnetization $\bm{M}$ and broken spatial inversion symmetry can have $\nu=0$: if the Fermi level lies in between the two Weyl points the circular Fermi surface is twisted into a \mbox{figure-8} with zero total curvature \cite{note0}. 

The self-intersection introduced when the Fermi level passes through a Weyl point, to ensure that $|\nu|+s$ remains odd, is a crossing of Fermi arcs on the top and bottom surfaces of the thin film (width $W$). These have a penetration depth $\xi_0$ into the thin film that can be much less than the Fermi wavelength of the bulk states, so that we can be in the 2D regime of a single occupied subband \cite{note3} without appreciable overlap of the surface states \cite{Pot14,Bul16a,Bul16b}. The effect of a nonzero surface state overlap is to open up an exponentially small gap $\delta k \propto e^{-W/\xi_0}$ in the figure-8, as in Fig.\ \ref{fig_turning}a.

In a perpendicular magnetic field $B$ the signed area enclosed by the Fermi surface is quantized in units of $2\pi/l_m^2$, with $l_m=\sqrt{\hbar/eB}$ the magnetic length. A \mbox{figure-8} Fermi surface of linear dimension $k_{\rm F}$ has a signed area much smaller than $k_{\rm F}^2$, because the upper and lower loops have opposite orientation. We find that this twisted Fermi surface produces edge states of width $k_{\rm F}l_m^2$ --- much wider than the usual narrow quantum Hall edge states of width $l_m$. The wide and the narrow edge states are counterpropagating: if the wide channel moves parallel to $\bm{M}$, the narrow channel moves antiparallel. An applied voltage selectively populates one of the two types of edge states, resulting in a conductance of $e^2/h$ instead of $2e^2/h$ --- even though there are two conducting edges.

The outline of the paper is as follows. In the next section we formulate the problem, on the basis of a two-band model Hamiltonian \cite{Yan11,Oga14}, and calculate the band structure in a slab geometry. The way in which the Fermi arcs reconnect with the bulk Weyl cones is described exactly by a simple transcendental equation (Weiss equation). The Fermi surface in the thin-film regime is calculated in Sec.\ \ref{thinfilmEF}, to show the topological transition from turning number 1 to turning number 0 when the Fermi level passes through a Weyl point. In Sec.\ \ref{QHEedge} we calculate the edge states in a perpendicular magnetic field, by semiclassical analytics and comparison with a numerical solution. The implications of the two types of counterpropagating edge channels for electrical conduction are investigated in Sec.\ \ref{transport}. We conclude with an overview of possible experimental signatures of the twisted Fermi surface.

\section{Weyl semimetal confined to a slab}
\label{minimalmodel}

\subsection{Two-band model}
\label{twobandmodel}

We consider the two-band model Hamiltonian of a Weyl semimetal \cite{Yan11,Oga14},
\begin{align}
&H(\bm{k})=t_x\sigma_x \sin k_x+t_y\sigma_y\sin k_y+m_{\bm k}\sigma_z+\lambda\sigma_0\sin k_z,\nonumber\\
&m_{\bm k}=t_z(\cos \beta-\cos k_z)+t'(2-\cos k_x-\cos k_y).\label{Hdef}
\end{align}
The Pauli matrices are $\sigma_\alpha$, $\alpha\in\{x,y,z\}$, with $\sigma_0$ the $2\times 2$ unit matrix, acting on a hybrid of spin and orbital degrees of freedom. The momentum $\bm{k}$ varies over the Brillouin zone $|k_\alpha| < \pi$ of a simple cubic lattice (lattice constant $a_0\equiv 1$, and we also set $\hbar\equiv 1$). The two Weyl points are at the momenta $\bm{k}=(0,0,\pm K)$, $K\approx\beta$, and at energies $E=\pm E_0$, $E_0\approx \lambda\sin\beta$, displaced along the $k_z$-axis by the magnetization $\bm{M}=\beta\hat{z}$ and displaced along the energy axis by the strain $\lambda$. Time-reversal symmetry and spatial inversion symmetry are broken by $\beta$ and $\lambda$, respectively.

We take a slab geometry, unbounded in the $y$--$z$ plane and confined in the $x$-direction between $x=0$ and $x=W$. The magnetization along $z$ is therefore in the plane of the slab. We impose the infinite-mass boundary condition \cite{Ber87} on the wave function $\psi$,
\begin{equation}
\sigma_y\psi=\begin{cases}
-\psi&\text{at}\;\;x=0,\\
+\psi&\text{at}\;\;x=W.
\end{cases}\label{infinitemassbc}
\end{equation}
This boundary condition corresponds to a mass term $m_0(x)\sigma_z$ in $H$ that vanishes inside the slab and tends to $+\infty$ outside.

\subsection{Dispersion relation}
\label{sec_dispersion}

The Schr\"{o}dinger equation $H\psi=E\psi$ can be solved analytically in the low-energy regime by linearizing in $k_x$ and applying the effective mass approximation \cite{Lut55} $k_x\mapsto -i\partial/\partial x$. Integration of the resulting first-order differential equation in $x$ gives
\begin{equation}
\psi(x)=e^{ix\Xi}\psi(0),\;\;
\Xi=\frac{1}{t_x}\sigma_x[E-H(0,k_y,k_z)].\label{Xidef}
\end{equation}
To ensure that an eigenstate of $H$ satisfies the boundary condition \eqref{infinitemassbc}, we require that
\begin{equation}
\langle -| e^{iW\Xi}|-\rangle=0,\;\;|\pm\rangle=\begin{pmatrix}
1\\
\pm i
\end{pmatrix},\;\;\sigma_y|\pm\rangle=\pm|\pm\rangle.
\end{equation}
This reduces to the following dispersion relation for $E(k_y,k_z)$:
\begin{equation}
(E-\lambda\sin k_z)^2-t_y^2 \sin^2 k_y-m_{\bm{k}}^2=q^2,\label{Eqdispersion}
\end{equation}
with transverse wave number $q$ given by
\begin{equation}
\frac{m_{\bm{k}}}{q}\tan(Wq/t_x)+1=0.\label{qkzkyrelation}
\end{equation}
In the mass term $m_{\bm{k}}$ we should set $k_x=0$, as required by the linearization in $k_x$.

For imaginary $q=i\kappa t_x/W$ the transcendental equation \eqref{qkzkyrelation} takes the form
\begin{equation}
\frac{\gamma}{\kappa}\tanh \kappa=1,\;\;\gamma=-\frac{Wm_{\bm{k}}}{t_x},\label{Weisseq}
\end{equation}
which is known as the Weiss equation in the theory of ferromagnetism \cite{Bar17}. A unique solution with $\kappa\geq 0$ exists for $\gamma\geq 1$, given by a generalized Lambert function \cite{Mez06,note1}:
\begin{equation}
\kappa=\tfrac{1}{2}{\cal W}(2\gamma;-2\gamma;-1).\label{Lamberteq}
\end{equation}

\begin{figure}[tb]
\centerline{\includegraphics[width=0.9\linewidth]{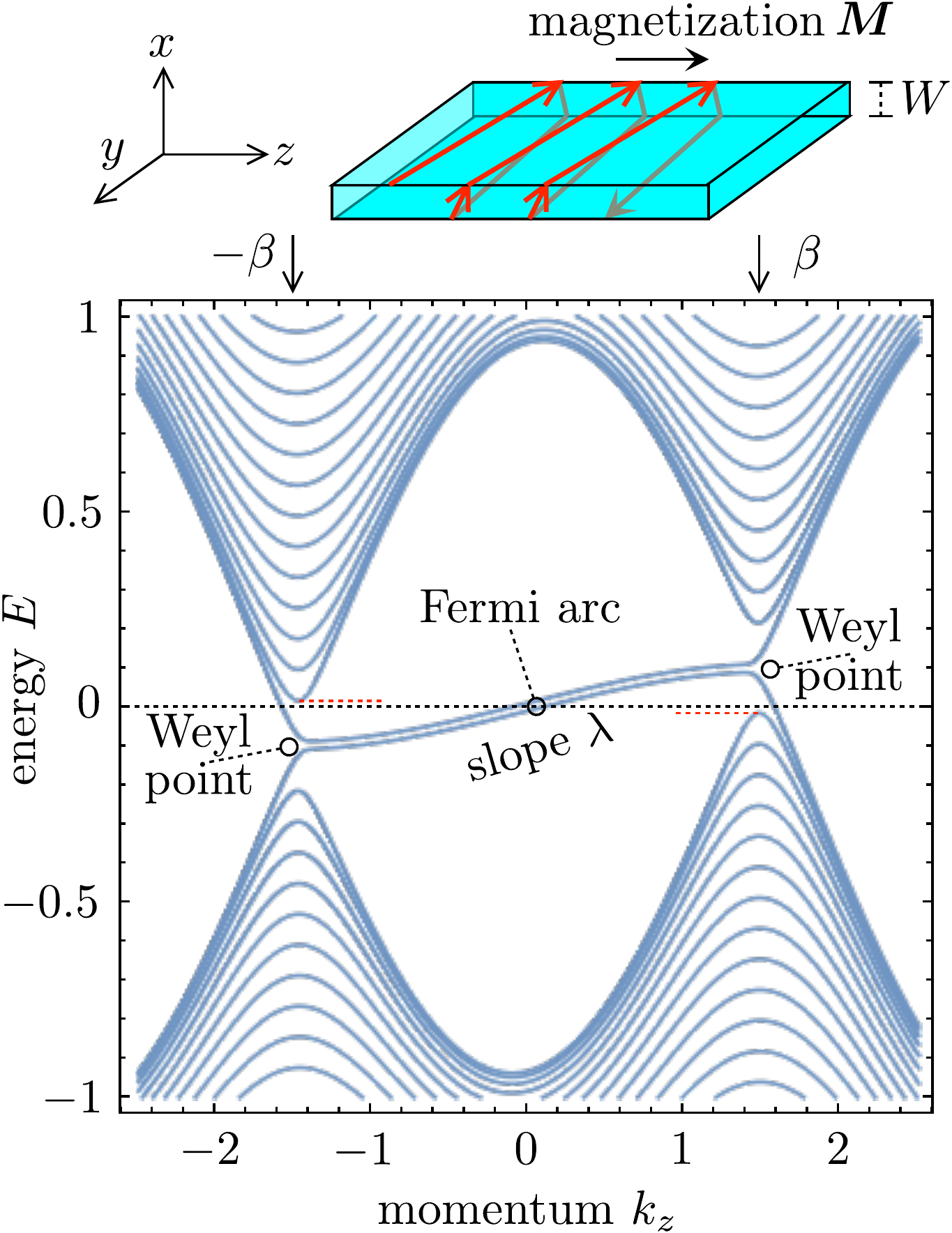}}
\caption{Dispersion relation $E(k_y,k_z)$ for $k_y=0.01$ as a function of $k_z$, of a thick Weyl semimetal slab (width $W=40$), calculated from Eqs.\ \eqref{Eqdispersion} and \eqref{qkzkyrelation} for $\beta=1.5$, $\lambda=0.1$, $t_x=t_y=t_z=t'=1$. The diagram at the top shows the geometry with the trajectory of an electron in a Fermi arc state spiralling along the surface with velocity $v_z=\lambda\cos k_z$ in the direction of the magnetization $\bm{M}$. The two branches of the Fermi arc visible in the dispersion relation correspond to states on the top and bottom surface of the slab (assumed to be of infinite extent in this calculation). For this thick slab the range of Fermi energies in which only a single 2D subband is occupied is very narrow (between the red dotted lines). For thinner slabs a larger energy range is available.
}
\label{fig_dispersion}
\end{figure}

A representative band structure is shown in Fig.\ \ref{fig_dispersion}.

\subsection{Weyl cones and Fermi arcs}

In the large-$W$ limit of a thick slab, Eq.\ \eqref{qkzkyrelation} can be solved separately for the bulk Weyl cones and the surface Fermi arcs. We thus recover the familiar dispersion relations in the bulk and surface Brillouin zones of a Weyl semimetal \cite{Has17,Yan17,Bur17,Arm17}.

The bulk states have wave number $q\gg |m_{\bm{k}}|$, quantized by $q=(n+\tfrac{1}{2})\pi t_x/W$, $n=0,1,2,\ldots$, with dispersion
\begin{align}
E_{\rm bulk}^{(n)}={}&\pm\sqrt{(n+\tfrac{1}{2})^2(\pi t_x/W)^2+t_y^2 \sin^2 k_y+m_{\bm{k}}^2}\nonumber\\
&+\lambda\sin k_z.\label{Ebulk}
\end{align}
The $\pm$ distinguishes the upper and lower halves of the Weyl cones.

The surface Fermi arcs have a purely imaginary $q=im_{\bm{k}}\Rightarrow\kappa=-\gamma$, which solves Eq.\ \eqref{Weisseq} in the large-$W$ limit if $m_{\bm{k}}<0$. The corresponding surface dispersion \eqref{Eqdispersion} is 
\begin{equation}
E_{\rm surface}=\lambda\sin k_z\pm t_y\sin k_y,\;\;|k_z|<\beta.\label{Esurface}
\end{equation}
The $\pm$ sign distinguishes the Fermi arcs on opposite surfaces ($-$ at $x=0$ and $+$ at $x=W$). The trajectory of an electron in a Fermi arc state moves chirally along the surface (see top inset in Fig.\ \ref{fig_dispersion}), spiralling in the direction of the magnetization $\bm{M}=\beta\hat{z}$ with velocity $v_z=\lambda \cos k_z$.

\begin{figure}[tb]
\centerline{\includegraphics[width=0.7\linewidth]{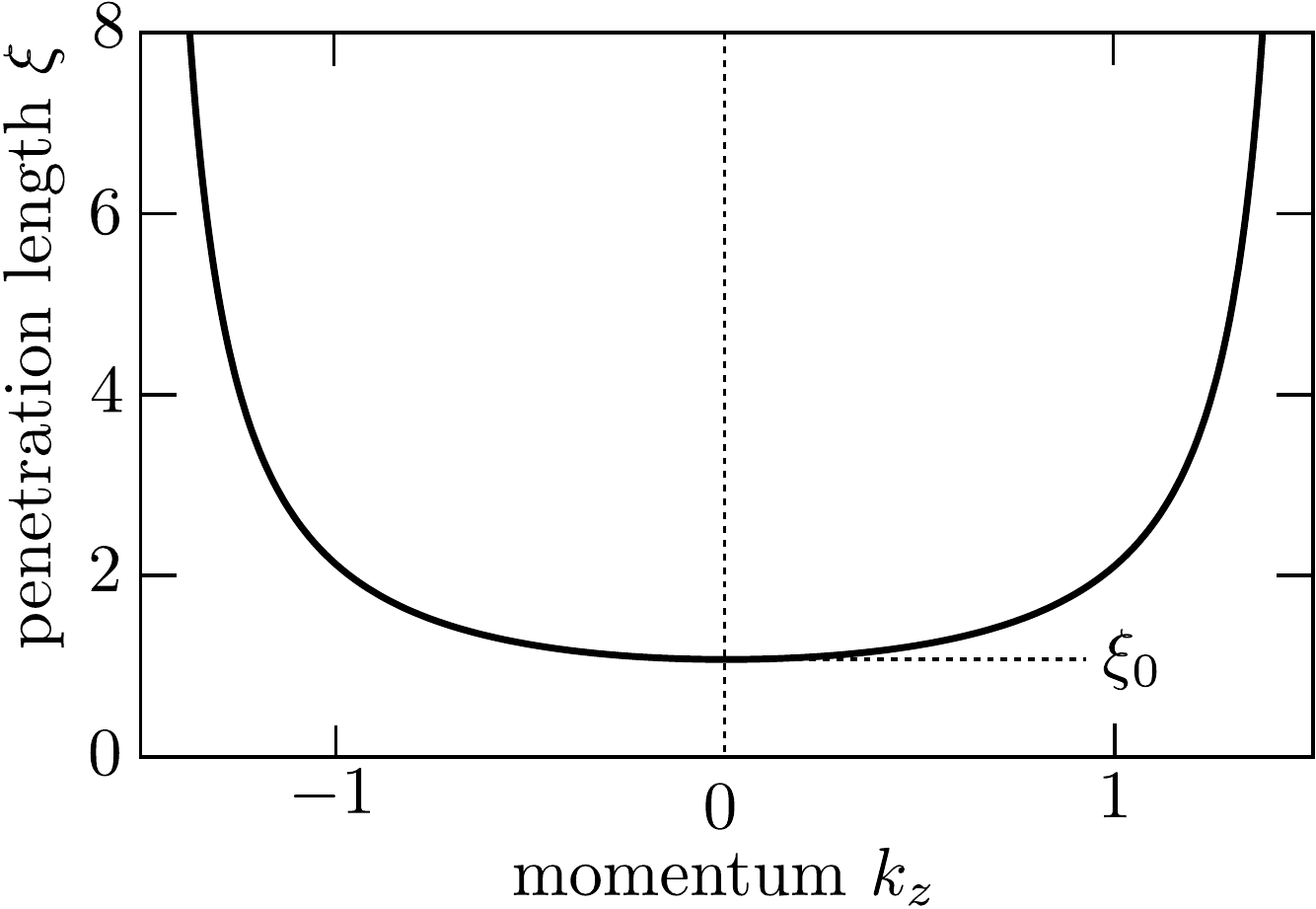}}
\caption{Penetration length $\xi$ of the surface Fermi arc into the bulk Weyl semimetal, calculated via $\xi=1/{\rm Im}\,q$ from the solution of the Weiss equation \eqref{Weisseq}, for the same parameters as Fig.\ \ref{fig_dispersion}. The penetration length diverges at $k_z=\pm 1.475$, according to Eq.\ \eqref{kzcrit}. At this critical momentum the Fermi arc merges with the bulk Weyl cones. The minimal penetration length $\xi_0$ is given by Eq.\ \eqref{xi0def}.
}
\label{fig_Weiss}
\end{figure}

The surface Fermi arc reconnects with the bulk Weyl cone near $k_z=\pm \beta$. This ``Fermi level plumbing'' \cite{Hal14} is described quantitatively by the Weiss equation \eqref{qkzkyrelation}, as $q$ switches from imaginary to real at a critical $k_z^{\rm crit}$ for which $\gamma=1$. The penetration length $\xi=1/{\rm Im}\,q$ of the surface state into the bulk is plotted in Fig.\ \ref{fig_Weiss}, as a function of $k_z$ for $k_y=0$. Its minimal value near the center of the Brillouin zone is
\begin{equation}
\xi_0=\frac{t_x}{(1-\cos\beta)t_z}.\label{xi0def}
\end{equation}
The critical wave vector $\bm{k}=(0,0,k_z^{\rm crit})$ at which the Fermi arc terminates because its penetration length diverges is slightly smaller than the position $\beta$ of the Weyl point,
\begin{equation}
k_z^{\rm crit}=\beta-\frac{t_x}{t_z W}+{\cal O}(W^{-2}).\label{kzcrit}
\end{equation}

\section{Thin-film Fermi surface}
\label{thinfilmEF}

\begin{figure}[tb]
\centerline{\includegraphics[width=1\linewidth]{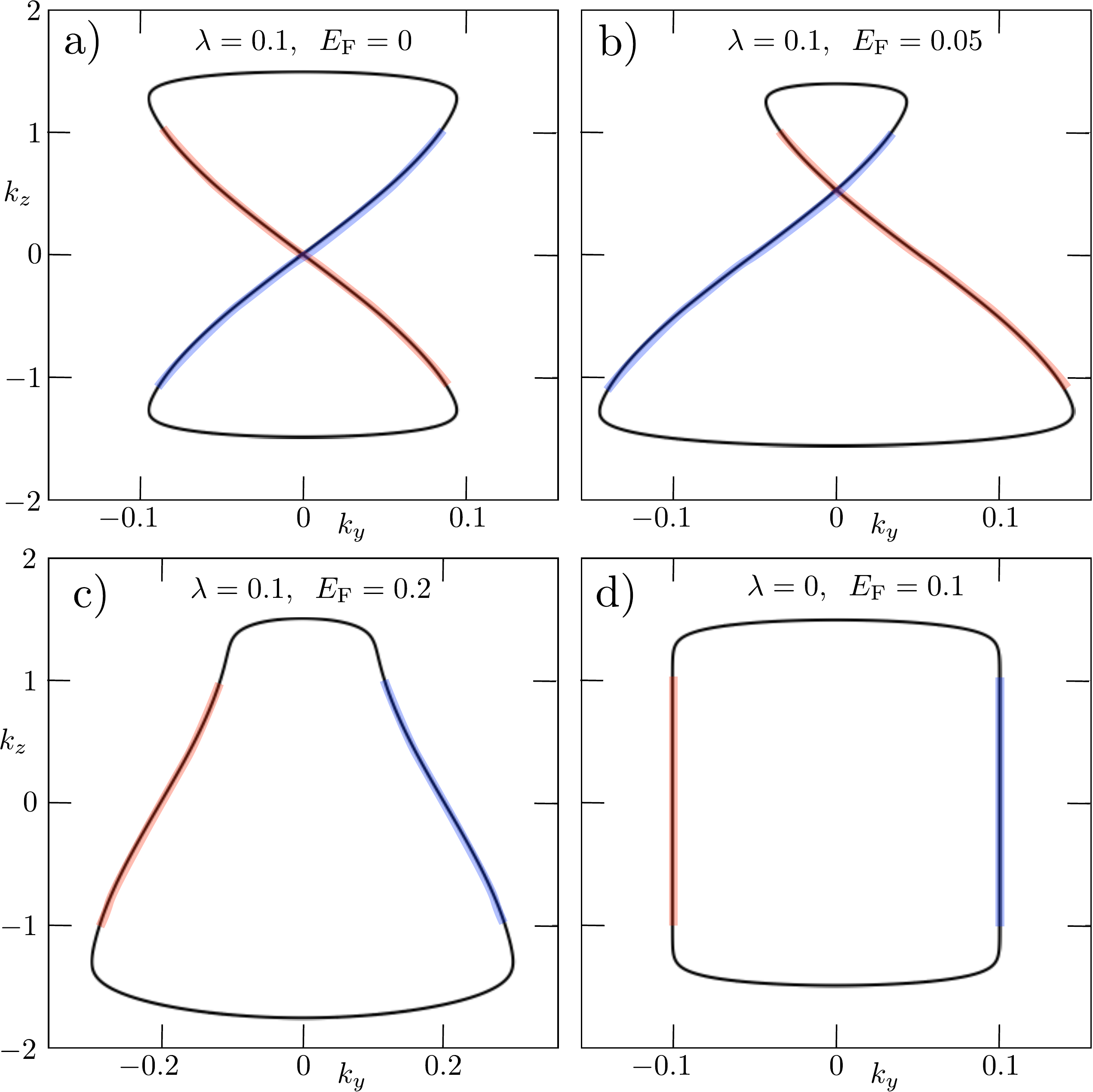}}
\caption{Fermi surfaces of the thin-film Weyl semimetal with a single occupied subband ($W=15$), calculated from Eqs.\ \eqref{Eqdispersion} and \eqref{qkzkyrelation} for $\beta=1.5$, $t_x=t_y=t_z=t'=1$ at different values of $\lambda$ and $E_{\rm F}$. The turning number $\nu=0$ in the top row, while $\nu=1$ in the bottom row. The \mbox{figure-8} in the top row has a narrowly avoided crossing with a gap $\delta k_z=3\cdot 10^{-5}$ (not visible on the scale of the figure). The color of the contour indicates whether the state is localized on the top surface (red), on the bottom surface (blue), or extended through the bulk (black).
}
\label{fig_Fermisurfaces}
\end{figure}

For Fermi energies 
\begin{equation}
|E_{\rm F}|<\frac{\pi t_x}{2W}-\lambda\sin\beta, \label{EFsingle}
\end{equation}
a single two-dimensional (2D) subband is occupied at the Fermi level, formed out of hybridized bulk and surface states. This two-dimensional electron gas (2DEG) regime exists for thin films of width
\begin{equation}
W\lesssim W_c=\frac{\pi t_x}{2\lambda\sin\beta}. \label{Wcdef}
\end{equation}
The Fermi surface of the 2DEG, defined by the equi-energy contour $E(k_y,k_z)=E_{\rm F}$, is plotted in Fig.\ \ref{fig_Fermisurfaces} for several parameter values.

As discussed in the introduction, the turning number $\nu$ is a topological invariant of the equi-energy contour \cite{homotopy}. We see from Fig.\ \ref{fig_Fermisurfaces} that the Fermi surface is twisted into a \mbox{figure-8} with $\nu=0$ when the Fermi level lies between the Weyl points, $|E_{\rm F}|<\lambda\sin\beta$, while for larger Fermi energies the Fermi surface has $\nu=1$. Because the turning number and the number of self-intersections must have opposite parity, the topological transition when $E_{\rm F}$ passes through a Weyl point must introduce a crossing in the Fermi surface \cite{Siu16}.

The crossing of the equi-energy contour for small $E_{\rm F}$ is possible since the intersecting states are spatially separated on the top and bottom surfaces of the slab. For a finite ratio $W/\xi_0$ of slab width and penetration length \eqref{xi0def} the crossing is narrowly avoided because of the exponentially small overlap of the states at opposite surfaces. From the Weiss equation \eqref{Weisseq} we calculate that the $\delta k_z$ gap in the \mbox{figure-8} is given by
\begin{equation}
\delta k_z=\frac{4t_x}{\lambda\xi_0}e^{-W/\xi_0}.\label{deltakz}
\end{equation}
When $W\simeq W_c$ the gap in the \mbox{figure-8} is exponentially small if $W_c\gg\xi_0$, so for
\begin{equation}
(1-\cos\beta)t_z\gg\lambda\sin\beta.\label{unbrokenfigureof8}
\end{equation}

To make contact with some of the older literature \cite{Zil58,Azb61,Rot66}, we note that the figure-8 Fermi surface of a Weyl semimetal is essentially different from the figure-8 equi-energy contour of a conventional metal with a saddle point in the Fermi surface. In that case the figure-8 requires fine tuning of the energy to the saddle point, while here the figure-8 persists over a range of energies between two Weyl points. Moreover, the orientation of the two lobes of the figure-8 is the same in the case of a saddle point, while here it is opposite.

\section{Quantum Hall edge channels}
\label{QHEedge}

\subsection{Semiclassical analysis}
\label{QHE_semiclassics}

A magnetic field $B$ in the $x$-direction, perpendicular to the thin film, introduces Landau levels in the energy spectrum: For a gauge $\bm{A}=(0,0,By)$ the momentum $k_z$ is still a good quantum number, we seek the dispersion $E_n(k_z)$ of the $n$-th Landau level.

Semiclassically, the $n$-the Landau level is determined by the quantization of the signed area $S(E)=\oint k_y dk_z$ enclosed by the oriented equi-energy contour \cite{Kos06}, 
\begin{equation}
l_m^2 S(E_n)=2\pi (n+\gamma),\;\;n\in\mathbb{Z},\label{semiclassics}
\end{equation}
with $l_m=(\hbar/eB)^{1/2}$ the magnetic length and $\gamma\in[0,1)$ a $B$-independent offset. Depending on the clockwise or anti-clockwise orientation of the contour, the enclosed area is negative or positive. Note that the signed area enclosed by the \mbox{figure-8} Fermi surface of Fig.\ \ref{fig_Fermisurfaces}a equals zero. The phase shift $\gamma=0$ in a bulk Weyl semimetal, when the equi-energy contour encloses a gapless Weyl point \cite{Mic99,Fuc10,Ale17,Ale17b}. For the thin film the numerical data indicates $\gamma=1/2$. 

If the thin film is confined to the strip $0<y<W_y$, with $W_y\gg l_m$, the spectrum within the strip remains dispersionless, but at the boundaries $y=0$ and $y=W_y$ propagating states appear. In the quantum Hall effect these are chiral edge channels, moving in opposite directions on opposite edges \cite{Hal82,But88}. The electrical conductance of the strip, for a current flowing in the $z$-direction, equals the number of edge channels $N$ moving in the same direction times the conductance quantum $e^2/h$.

\begin{figure}[tb]
\centerline{\includegraphics[width=0.9\linewidth]{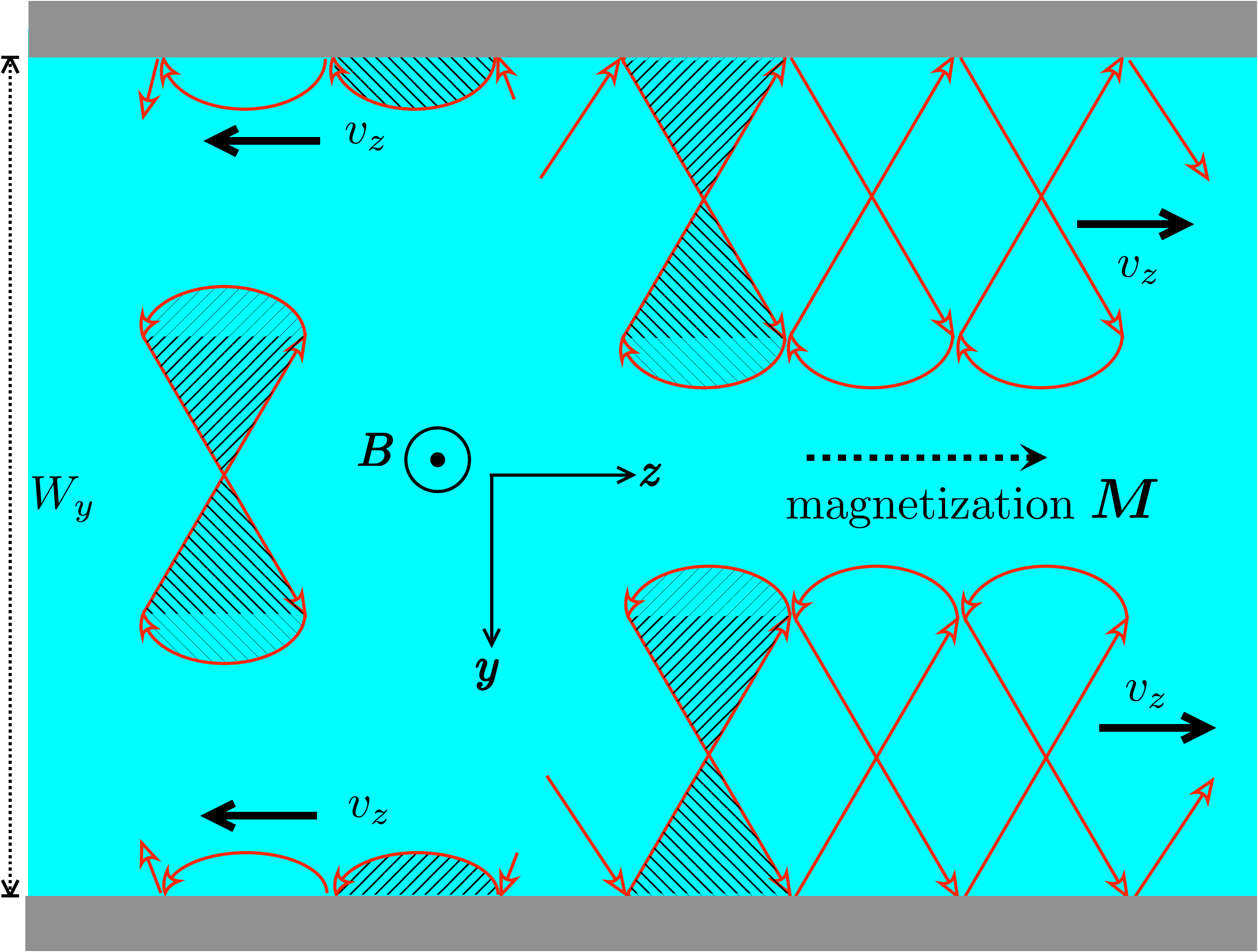}}
\caption{Classical cyclotron orbits corresponding to the \mbox{figure-8} Fermi surface of Fig.\ \ref{fig_Fermisurfaces}a. Each edge supports counterpropagating skipping orbits. The corresponding quantum Hall edge channel is narrow if it propagates opposite to the magnetization, while it is wide if it propagates in the direction of the magnetization. The area enclosed by the cyclotron orbits is shaded, the direction of the shading distinguishes positive and negative contributions to the Aharonov-Bohm phase $e\oint \bm{A}\cdot d\bm{l}$.
}
\label{fig_edgestates}
\end{figure}

The classical skipping orbits that form the edge channels in a magnetic field can be directly extracted from the zero-field Fermi surface: The cyclotron motion in momentum space follows the equi-energy contour $E(k_y,k_z)=E_{\rm F}$ with period $2\pi m_c/eB$, where
\begin{equation}
m_{c}=\frac{1}{2\pi}\frac{d}{dE}|S(E)|\label{mcdef}
\end{equation}
is the cyclotron effective mass. (The figure-8 has $m_c\approx\beta/t_y$.) Because $\dot{\bm{k}}=e\dot{\bm{r}}\times\bm{B}$, the cyclotron motion in real space is obtained from the momentum space orbit by rotation over $\pi/2$ and rescaling by a factor $l_m^2$. Specular reflection at the edge (with conservation of $k_z$) then gives for the \mbox{figure-8} Fermi surface the skipping orbits of Fig.\ \ref{fig_edgestates}. Note that these orbits are 2D projections of 3D trajectories in the thin film:  The intersections that are visible in the projected orbit correspond to overpassing trajectories on the top and bottom surfaces. (See Fig.\ 10(b) of Ref.\ \onlinecite{Yao17} for a wave packet simulation of such a trajectory.)

The real-space counterpart of the quantization rule \eqref{semiclassics} is that the Aharonov-Bohm phase $e\oint \bm{A}\cdot d\bm{l}$ picked up in one period of the cyclotron motion equals $2\pi(n+\gamma)$. For the skipping orbits this Bohr-Sommerfeld quantization rule still applies if the contour is closed by a segment along the edge, with an additional contribution to $\gamma$ from reflection at the edge \cite{Bee89,Mon11}. 

For small $n$ the skipping orbit should enclose a flux of the order of the flux quantum $h/e$, which divides the edge channels into two types, designated narrow and wide: The narrow edge channel propagates along the edge in the direction opposite to the magnetization \cite{note2}. It is tightly bound to the edge over a distance of order $l_m$, so that the enclosed area of order $l_m^2$ encloses a flux of order $h/e$. The wide edge channel propagates in the direction of the magnetization and extends further from the edge over a distance of order $\beta l_m^2$. It still encloses a small flux of order $h/e$ because contributions to $\oint \bm{A}\cdot d\bm{l}$ from the two sides of the crossing point have opposite sign.

The gap $\delta k_z$ at the crossing point has no effect on the quantization if $l_m\delta k_z\ll 1$, which is satisfied for $l_m\lesssim W$ when
\begin{equation}
(W/\xi_0)e^{-W/\xi_0}\ll \lambda/t_x.\label{crossingpointtunneling}
\end{equation}
Because the exponent wins it is sufficient that $W\gg\xi_0$ to ensure that the \mbox{figure-8} is effectively unbroken: The field-induced tunneling through the gap then occurs with near-unit probability, so to a good approximation the wave packet propagates in an unbroken figure-8.

\begin{figure}[tb]
\centerline{\includegraphics[width=0.9\linewidth]{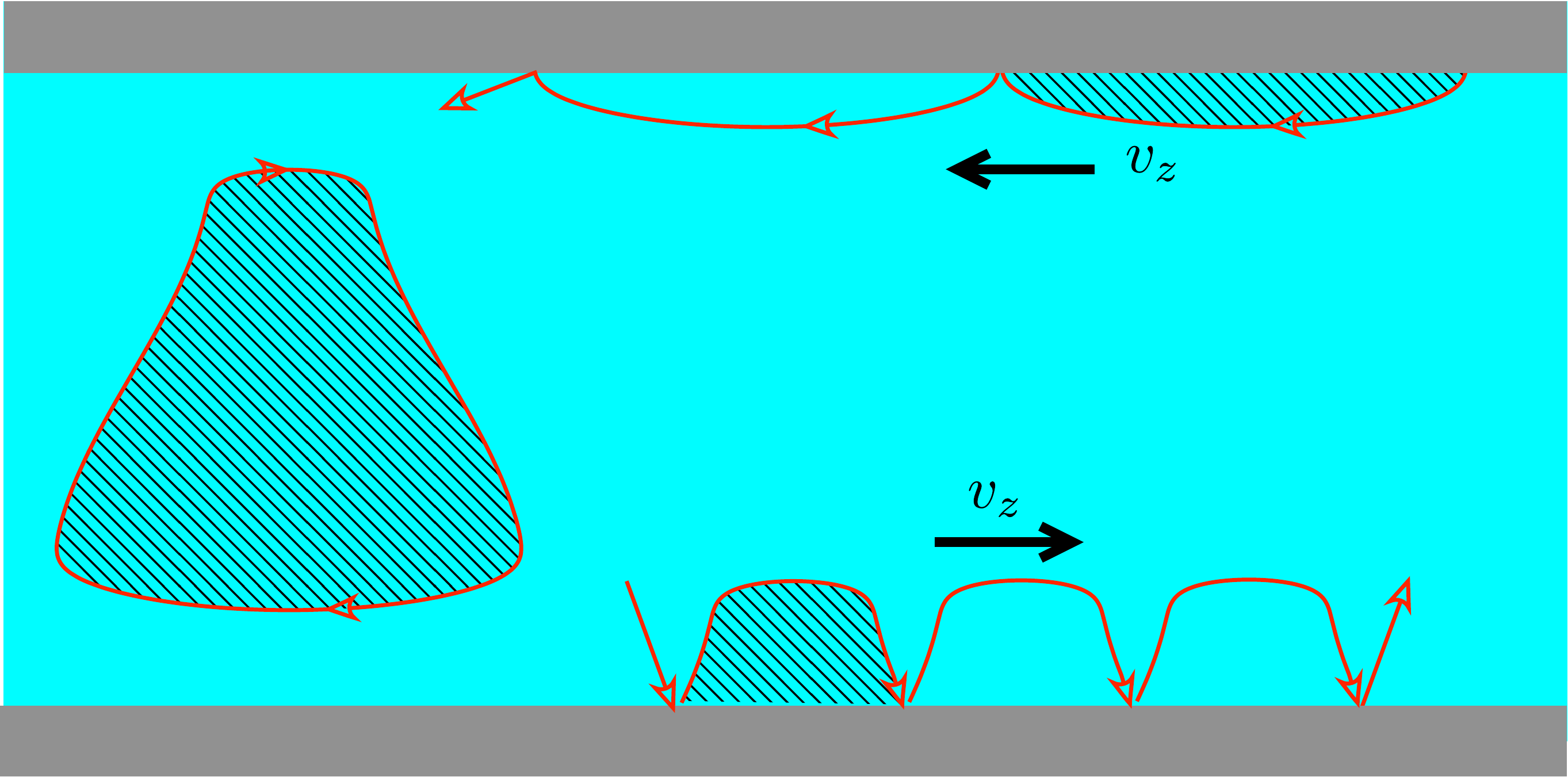}}
\caption{Same as Fig.\ \ref{fig_edgestates}, but now for the Fermi surface of Fig.\ \ref{fig_Fermisurfaces}c, without a self-intersection. The equi-energy contour has a single orientation, indicated by the single direction of the shading. The edge states are chiral, propagating in opposite directions on opposite edges.
}
\label{fig_chiral}
\end{figure}

The presence of counterpropagating edge channels at each edge requires a Fermi energy in between the Weyl points, $|E_{\rm F}|<\lambda\sin\beta$, for a twisted Fermi surface. When the Fermi surface is a simple contour without self-intersections the edge channels are chiral, propagating in opposite directions on opposite edges as in Fig.\ \ref{fig_chiral}.

\subsection{Numerical simulation}
\label{QHE_numerics}

\begin{figure}[tb]
\centerline{\includegraphics[width=0.8\linewidth]{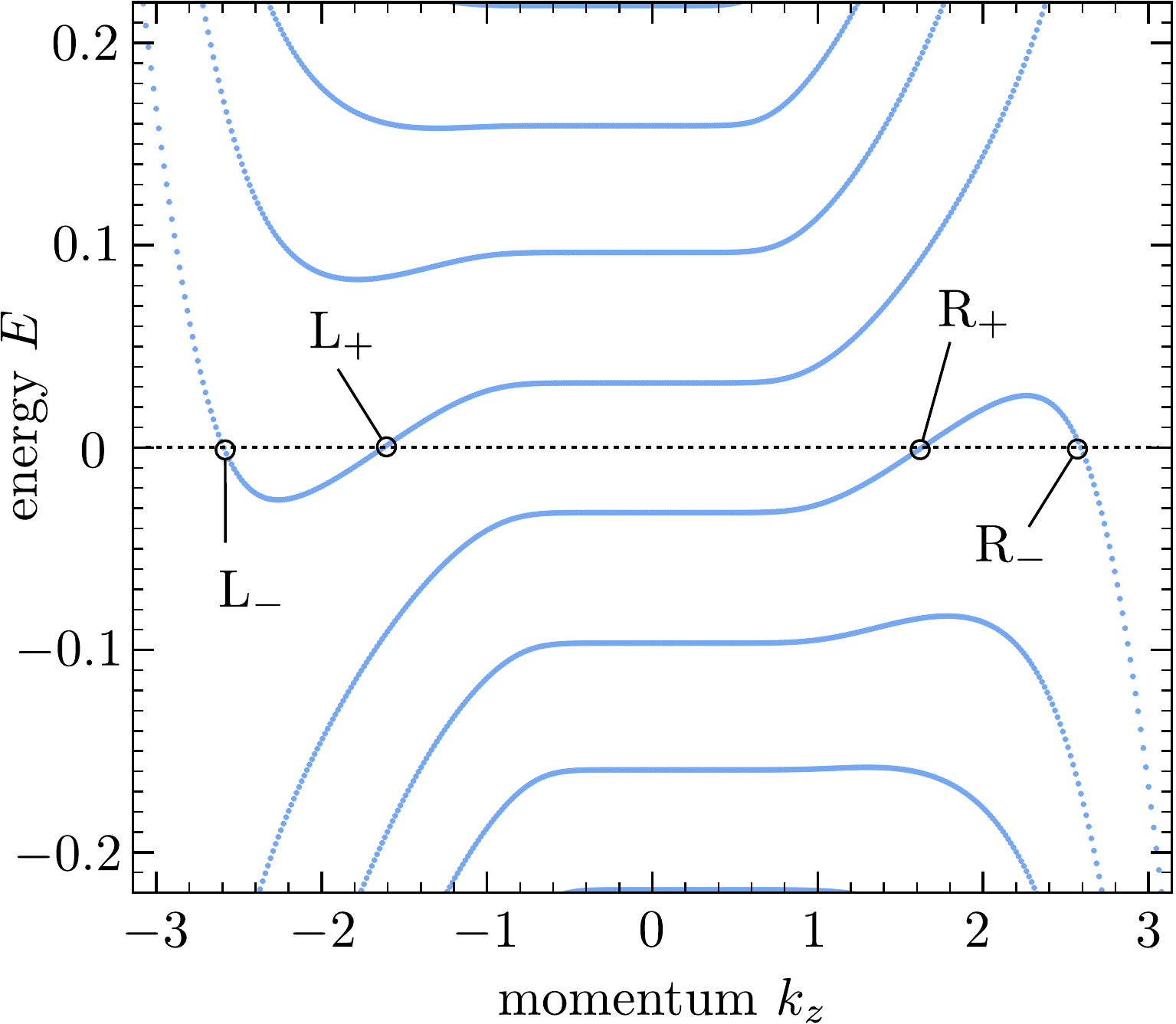}}
\caption{Dispersion relation of a thin-film Weyl semimetal strip ($W=10$, $W_y=80$) in a perpendicular magnetic field ($l_m=4.5$), calculated numerically from the tight-binding Hamiltonian \eqref{Hdef}. The material parameters are $\beta=1.05$, $\lambda=0.2$, $t_x=t_y=t_z=t'=1$. At $E_{\rm F}=0$ this system has the \mbox{figure-8} Fermi surface of Fig.\ \ref{fig_Fermisurfaces}a. The letters indicate the counterpropagating edge channels, ${\rm L}_\pm$ at one edge and ${\rm R}_\pm$ at the opposite edge.
}
\label{fig_numdisp}
\end{figure}

\begin{figure}[tb]
\centerline{\includegraphics[width=0.9\linewidth]{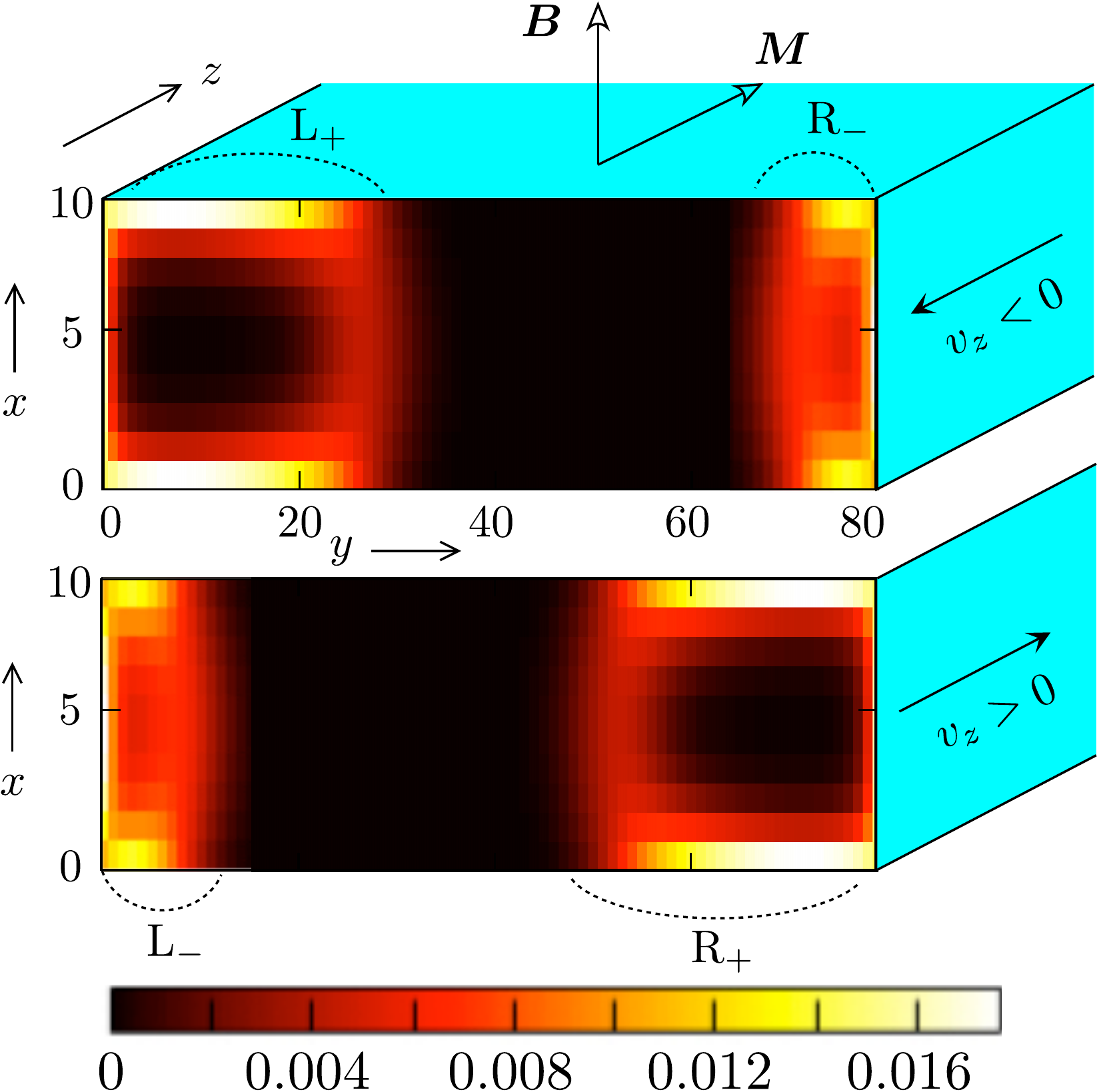}}
\caption{Probability density $|\psi(x,y)|^2$ for the four edge states labeled in the dispersion of Fig.\ \ref{fig_numdisp}. The density is translationally invariant in the $z$-direction, the color plots show a cross section in the $x$--$y$ plane (separated in two panels for clarity). Each edge has a counterpropagating pair of edge states, one with $v_z<0$ tightly bound to the edge (width $\approx l_m=4.5$), the other with $v_z>0$ penetrating more deeply into the bulk (width $\approx \beta l_m^2=21$).
}
\label{fig_edgeprofile}
\end{figure}

To go beyond the semiclassical analysis we have diagonalized the model Hamiltonian \eqref{Hdef} numerically, using the \textit{Kwant} tight-binding code \cite{kwant}. Fig.\ \ref{fig_numdisp} shows the dispersion relation with four edge states at $E_{\rm F}=0$, two counterpropagating at each edge. The corresponding density profile for each edge state is shown in Fig.\ \ref{fig_edgeprofile}. The two types of edge channels, one wide and the other narrow, are clearly visible.

\begin{figure}[tb]
\centerline{\includegraphics[width=1\linewidth]{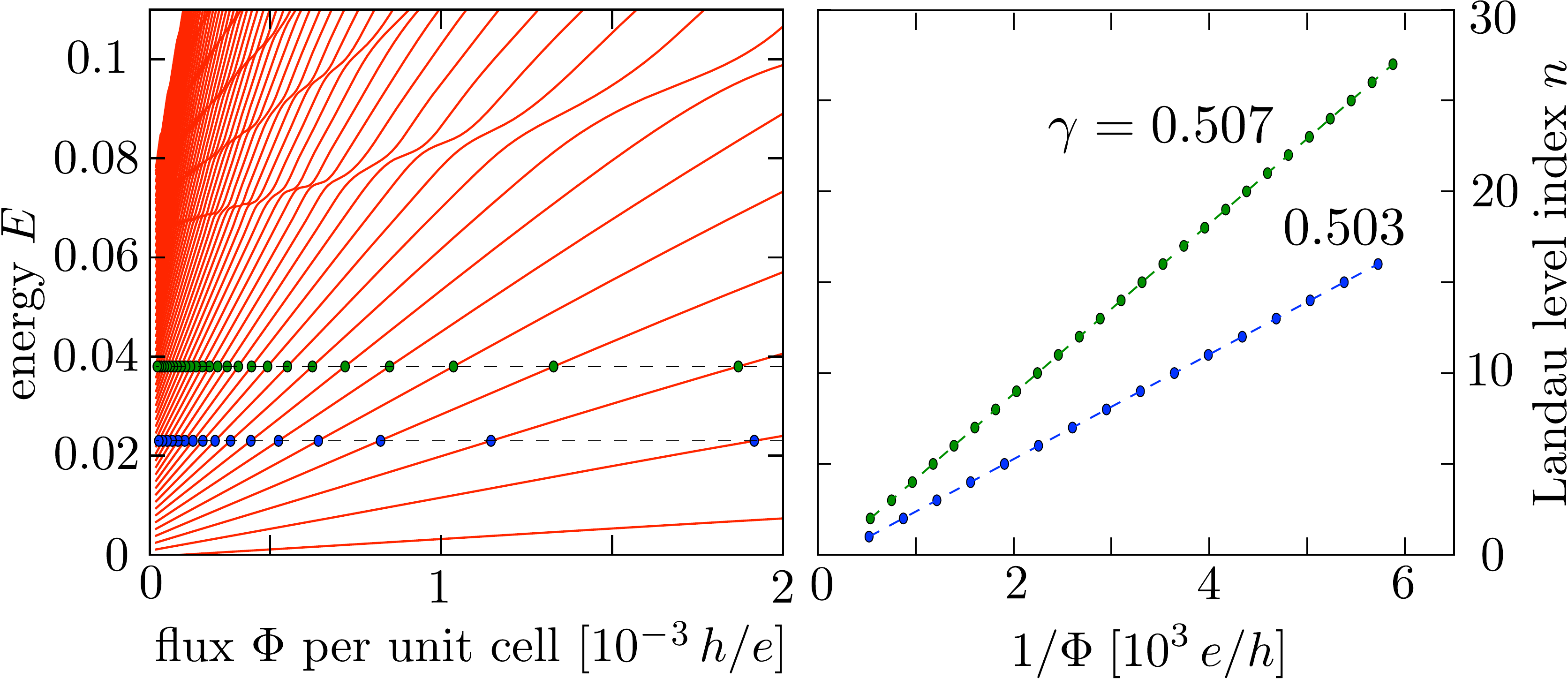}}
\caption{Left panel: Sequence of Landau level energies $E_n(B)$ as a function of magnetic field; levels at two values of the energy are marked by colored dots. Right panel: Landau level index $n$ for these two energies as a function of inverse magnetic field. This ``Landau fan'' is fitted to Eq.\ \eqref{fanfit} to obtain the offset $\gamma$. The data is calculated numerically from the Weyl semimetal tight-binding model in an unbounded thin film (thickness $W=30$), for parameters $\beta=1.05$, $\lambda=0.1$, $t_x=t_y=t_z=t'=1$.
}
\label{fig_landaufan}
\end{figure}

In Fig.\ \ref{fig_landaufan} we show the Landau levels in an infinite system as a function of the flux $\Phi$ through a unit cell. The Landau fan is fitted to
\begin{equation}
\frac{\hbar}{e\Phi}S_{E}=2\pi(n+\gamma),\label{fanfit}
\end{equation}
corresponding to the semiclassical formula \eqref{semiclassics}. The resulting offset $\gamma$ is consistent with $\gamma=1/2$. We checked that the fitted value of $S_E$ is close (within 2\%) of the signed area enclosed by the figure-8 equienergy contour. We also checked that the same $\gamma=1/2$ is obtained when the equienergy contour is a slightly deformed circle, rather than a figure-8.

\section{Magnetoconductance}
\label{transport}

To determine the magnetotransport through the Weyl semimetal strip we connect it at both ends $z=0$ and $z=L$ to a metal reservoir. Following a similar approach used for graphene \cite{Two06}, it is convenient to take the same model Hamiltonian \eqref{Hdef} throughout the system, with the addition of a $z$-dependent chemical potential term $-\mu(z)\sigma_0$. (Physically, this potential could be controlled by a gate voltage.) We set $\mu(z)=0$ in the semimetal region $0<z<L$ and take $\mu(z)\gg E_0$ in the metal reservoirs ($x<0$ and $x>L$). This corresponds to n-type doping of the reservoir. (For p-type doping we would take $\mu(z)\ll-E_0$.)

\begin{figure}[tb]
\centerline{\includegraphics[width=1\linewidth]{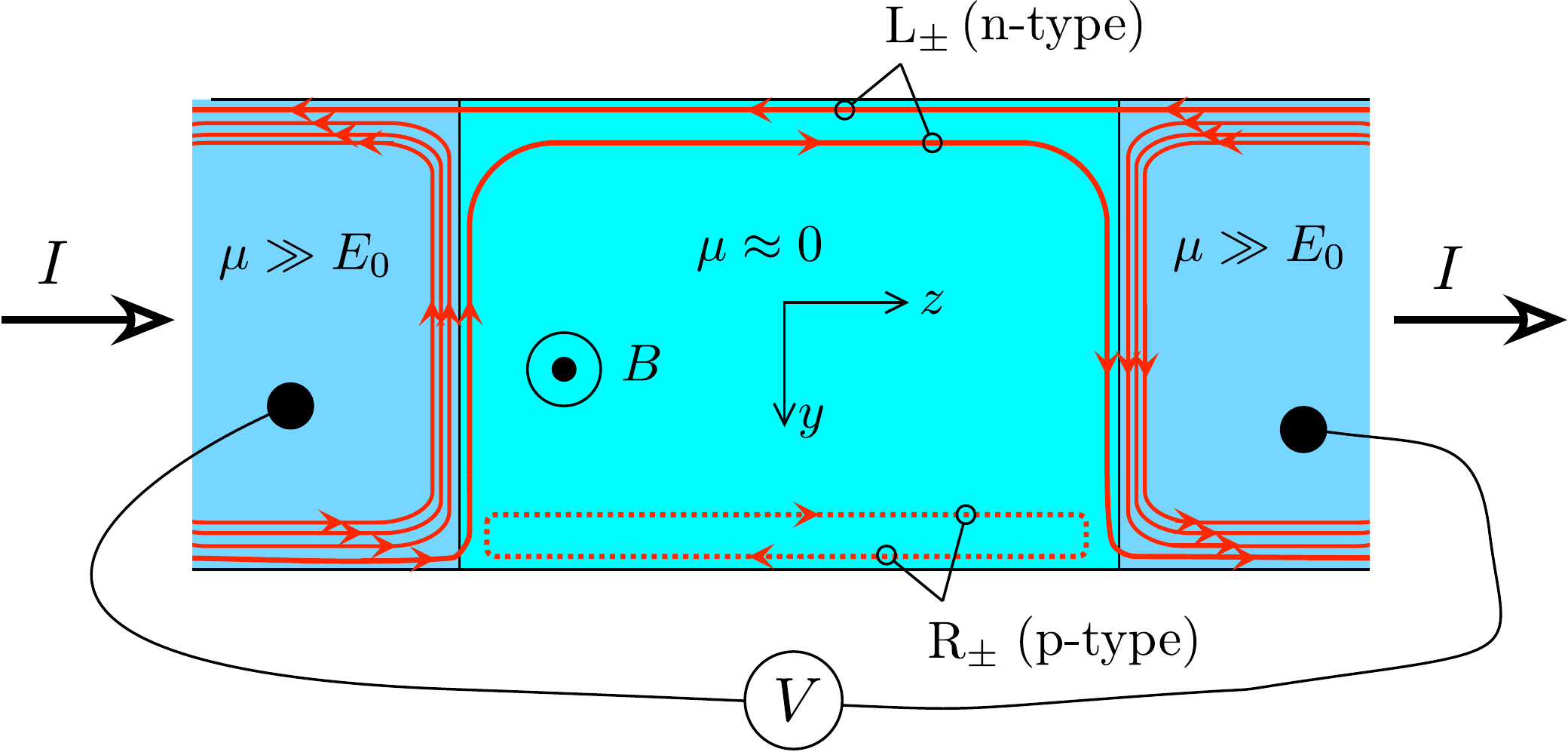}}
\caption{Undoped Weyl semimetal (chemical potential $\mu\approx 0$) connected to heavily doped metal reservoirs ($\mu\gg E_0$ for n-type doping). Edge channels in a perpendicular magnetic field are shown in red, with arrows indicating the direction of propagation. The ${\rm L}_\pm$ edge channels are n-type and can enter into the reservoirs, while the ${\rm R}_\pm$ edge channels are p-type and remain confined to the semimetal region (dotted lines). The current $I$ flows along the n-type edge in the semimetal, irrespective of the sign of the applied voltage $V$.
}
\label{fig_twoterminal}
\end{figure}

We distinguish n-type and p-type edge channels in the Weyl semimetal depending on whether they reconnect at large $|E|$ with the upper Weyl cones (n-type) or with the lower Weyl cones (p-type). Referring to the dispersion of Fig.\ \ref{fig_numdisp}, the channels ${\rm L}_\pm$ at the $y=0$ edge are n-type, while the channels ${\rm R}_\pm$ at the $y=W_y$ edge are p-type. The distinction is important, because only the n-type edge channels can be transmitted into the n-type reservoirs. As indicated in Fig.\ \ref{fig_twoterminal}, the p-type channels are confined to the semimetal region, without entering into the reservoirs.

\begin{figure}[tb]
\centerline{\includegraphics[width=0.7\linewidth]{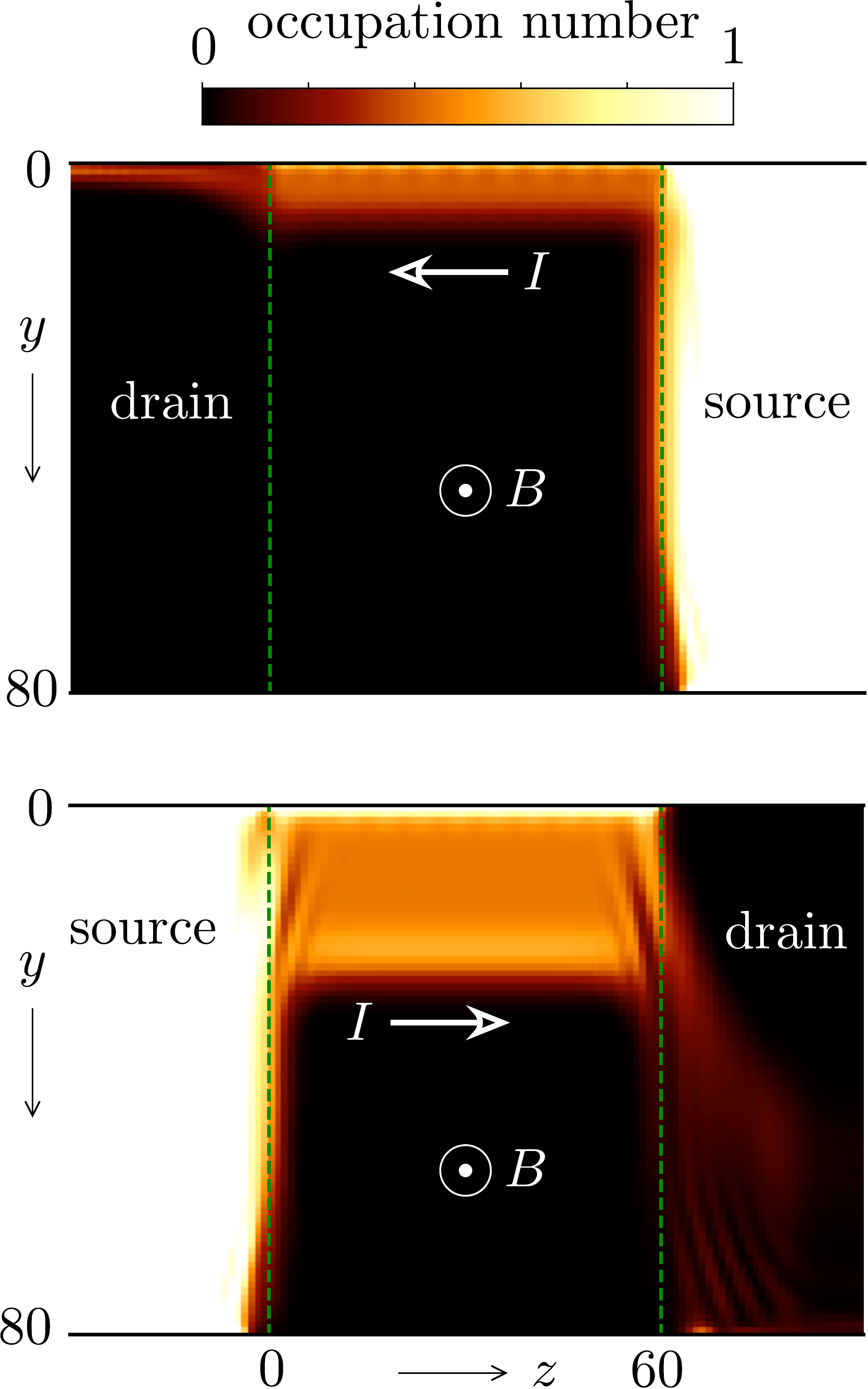}}
\caption{Color-scale plot in the $y$--$z$ plane of the occupation numbers of current-carrying states at the Fermi level, in response to a voltage bias between source and drain. The data is calculated numerically from the tight-binding Hamiltonian \eqref{Hdef} in the geometry of Fig.\ \ref{fig_twoterminal} (parameters $\beta=1.05$, $\lambda=0.25$, $t_x=t_y=t_z=t'=1$, $W=10$, $l_m=4$). The chemical potential is $\mu=0$ in the Weyl semimetal region (between green lines, from $z=0$ to $z=60$), while $\mu=0.75$ in the metal reservoirs ($z<0$ and $z>60$). The current keeps flowing along the same edge when source and drain are switched, carried either by a narrow edge channel (top panel) or by a wide edge channel (bottom panel). The opposite edge is fully decoupled from the reservoirs.
}
\label{fig_current}
\end{figure}

\begin{figure}[tb]
\centerline{\includegraphics[width=0.8\linewidth]{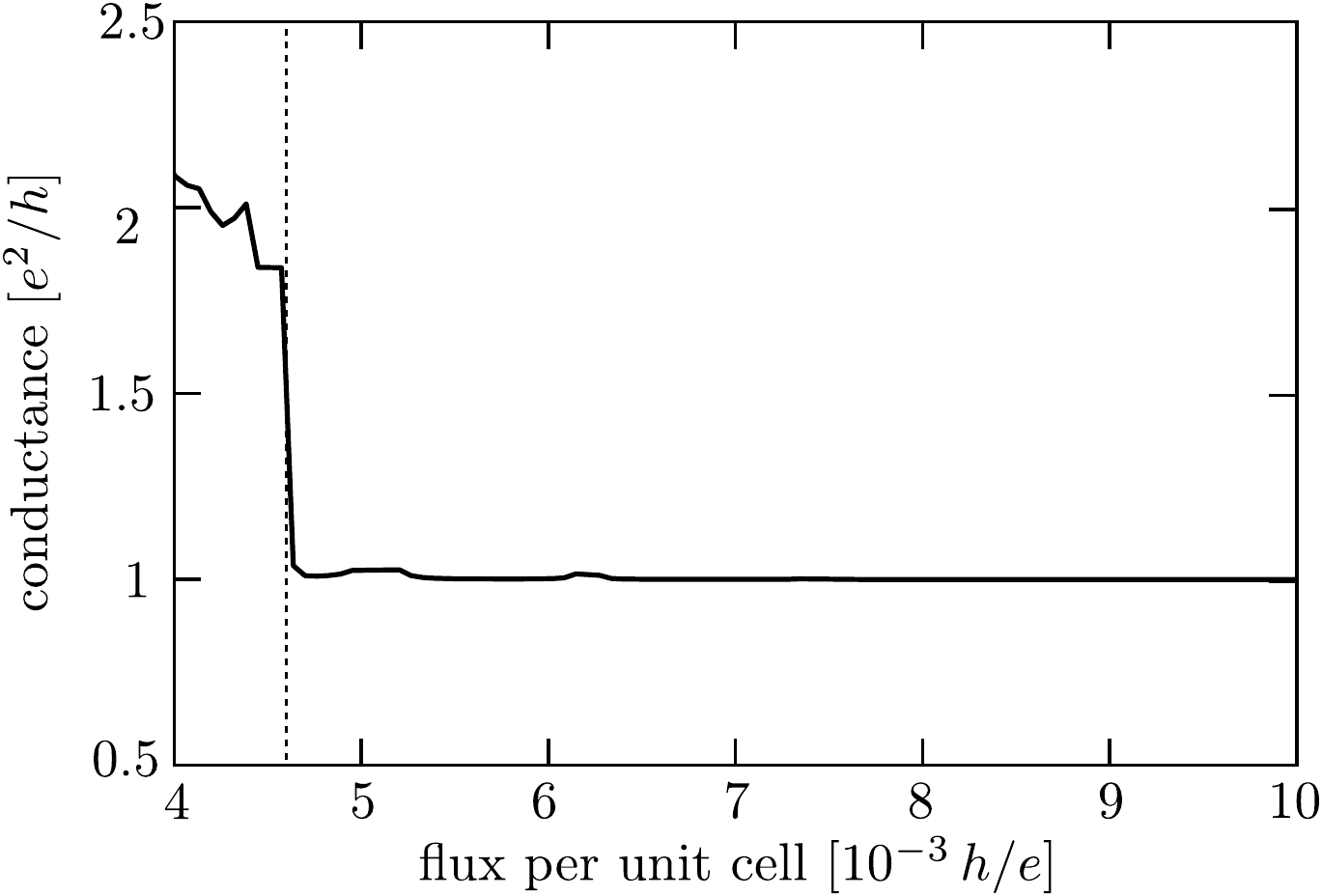}}
\caption{Conductance in the geometry of Fig.\ \ref{fig_current} as a function of magnetic field. (The magnetic length $l_m=4$ of Fig.\ \ref{fig_current} corresponds to a flux per unit cell of $0.01\,h/e$.) The regime of a single pair of counterpropagating edge channels is reached to the right of the vertical dotted line. The conductance in this regime is $e^2/h$ rather than $2e^2/h$, because only one edge is coupled to the electron reservoirs.
}
\label{fig_conductance}
\end{figure}

Upon application of a bias voltage $V$ between the two n-type reservoirs a current $I$ will flow along the n-type edge, with a conductance
\begin{equation}
G=I/V=\frac{e^2}{h}T_{y=0}\label{GTn}
\end{equation}
determined by the backscattering probability $T_{y=0}$ along the edge at $y=0$, so $G=e^2/h$ without impurity scattering --- see Fig.\ \ref{fig_conductance}.  This is not the usual edge conduction of the quantum Hall effect: As shown in Fig.\ \ref{fig_current}, the current flows along the \textit{same} edge when we change the sign of the voltage bias (switching source and drain), while in the quantum Hall effect the current switches between the edges when $V$ changes sign. The only way to switch the edge here is to change the sign of the magnetic field, so that the n-type edge is at $y=W_y$ rather than at $y=0$.

\section{Discussion}
\label{discuss}

We have discussed the unusual magnetic response of a two-dimensional electron gas with a twisted Fermi surface. The topological transition from turning number $\nu=1$ (the usual deformed Fermi circle) to turning number $\nu=0$ (the figure-8 Fermi surface) happens when the Fermi level passes through the Weyl point of a thin-film Weyl semimetal with an in-plane magnetization and broken spatial inversion symmetry. We discuss several transport properties that could serve as signatures for the topological transition from $\nu=1$ to $\nu=0$.

In a magnetic field the figure-8 Fermi surface supports counterpropagating edge channels, see Fig.\ \ref{fig_twoterminal}. At $E_{\rm F}=0$, with an equal number of left-movers and right-movers at each edge, the Hall resistance will vanish. This is the first magnetotransport signature. If we vary the Fermi level and enter the regime of chiral edge channels, we should see the appearance of a voltage difference between the edges in response to a current flowing along the edges. 

The second signature is the edge-selectivity: although both edges support counterpropagating states, the current flows entirely along one of the two edges, determined by the direction of $\bm{M}\times\bm{B}$. This edge-selective current flow might be detected directly, or indirectly by introducing disorder on one edge only and measuring a difference between the conductance $G$ for positive and negative $B$. Note that $G(B)\neq G(-B)$ does not violate Onsager reciprocity, since for that we would need to change the sign of both magnetic field $\bm{B}$ and magnetization $\bm{M}$.

A third signature is in the cyclotron resonance condition for the optical conductivity $\bm{\sigma}$. As explained by Koshino \cite{Kos16} in the context of a type-II Weyl semimetal (which has a figure-8 cyclotron orbit at a specific energy where electron and hole pockets touch \cite{OBr16}), the resonance frequency is twice as small for an electric field oriented along the long axis of the figure-8, than it is for an electric field oriented along the short axis. In the geometry of Fig.\ \ref{fig_edgestates}, the resonance frequency equals $eB/m_c$ for $\sigma_{yy}$ and $2eB/m_c$ for $\sigma_{zz}$.

In our analysis we have not included disorder effects. The counterpropagating edge channels can be coupled by disorder, and this would reduce the conductance below the quantized value of $G=e^2/h$ seen in Fig.\ \ref{fig_conductance}. There is no symmetry to protect this quantization, like there is for the helical edge channels in the quantum spin Hall effect, but there is a spatial separation of wide and narrow edge channels (see Fig.\ \ref{fig_edgeprofile}), which may provide some robustness against backscattering by disorder.

We have focused here on Fermi surfaces with turning number $\nu=0$ and $\nu=1$. It would be of interest to compare with other values of $\nu$. A model Hamiltonian for $\nu=2$, that could be a starting point for such a study, is given in the Appendix.

\acknowledgments
We have benefited from discussions with Hridis Pal. This research was supported by the Netherlands Organization for Scientific Research (NWO/OCW) and an ERC Synergy Grant.

\appendix

\section{Effective 2D Hamiltonian}
\label{app}

We derive an effective
Hamiltonian for the thin-film Weyl semimetal. Starting from the full Hamiltonian \eqref{Hdef}, we discretize the $x$-direction by the substitution
\begin{equation}
\begin{split}
&\cos k_x\mapsto \tfrac{1}{2}\big(\delta_{i,j-1}+\delta_{i,j+1}\big),\\
&\sin k_x\mapsto -\tfrac{1}{2}i\big(\delta_{i,j-1}-\delta_{i,j+1}\big).
\end{split}
\end{equation}
The Kronecker $\delta_{ij}$ is set to zero if either layer index $i$ or $j$ is outside of the set $\{1,2,\ldots, W\}$, corresponding to hard-wall boundary conditions at the top and bottom layer. Substitution in Eq.\ \eqref{Hdef} leads to
\begin{align}
H_{ij} ={}& \delta_{ij}\big[ \sigma_y\sin k_y
+M_{\bm k}\sigma_z\big] 
- \tfrac{1}{2}\delta_{i,j-1} \big(\sigma_z+i\sigma_x\big) \nonumber \\
&{} - \tfrac{1}{2}\delta_{i,j+1} \big(\sigma_z-i\sigma_x\big)+\delta_{ij}\lambda\, \sigma_0\sin k_z,\label{Hij1}\\
M_{\bm k}={}&2+\cos\beta-\cos k_z-\cos k_y.
\end{align}
For simplicity we have set 
$t_x=t_y=t'\equiv 1$. Since the $\lambda$ term is a scalar, we can set it to zero for now and then add it at the end of the calculation.

After the unitary transformation $H\mapsto U^\dagger HU$ with $U=e^{i\pi\sigma_z/4}e^{i\pi\sigma_y/4}$ we have
\begin{align}
H_{ij} ={}& \delta_{ij}\big[ \sigma_z\sin k_y
+M_{\bm k}\sigma_x\big] 
- \tfrac{1}{2}\delta_{i,j-1} \big(\sigma_x+i\sigma_y\big) \nonumber \\
&{} - \tfrac{1}{2}\delta_{i,j+1} \big(\sigma_x-i\sigma_y\big).\label{Hij2}
\end{align}
The square $H^2$ is block-diagonal 
in the $\sigma$ index,
\begin{subequations}
\begin{align}
&(H^2)_{ij}=\delta_{ij}\sigma_0\sin^2 k_y+\begin{pmatrix}
Z_{ij}&0\\
0&Z'_{ij}
\end{pmatrix},\\
&Z_{ij}=
 (M_{\bm k}^2+1-\delta_{iW})
\delta_{ij}-M_{\bm k}(\delta_{i,j-1}+\delta_{i,j+1}),\\
&Z'_{ij}=
 (M_{\bm k}^2+1-\delta_{i1})
\delta_{ij}-M_{\bm k}(\delta_{i,j-1}+\delta_{i,j+1}).
\end{align}
\end{subequations}

The two $W\times W$ matrices $Z$ and $Z'$ have the same eigenvalues $\zeta$, given by
\begin{equation}
{\rm Det}\,(Z-\zeta)=({\rm Det}\,Z)\left[1-\zeta\,{\rm Tr}\,Z^{-1}+{\cal O}(\zeta^2)\right]=0.
\end{equation}
The low-energy spectrum is therefore given
\begin{equation}
E^2=\sin^2 k_y +\zeta_0,\;\;\zeta_0=\frac{1}{{\rm Tr}\,Z^{-1}}\ll 1,
\end{equation}
which evaluates to
\begin{align}
\zeta_0 ={}& \frac{M_{\bm k}^{2W}}{1+2M_{\bm k}^2+3M_{\bm k}^4+4M_{\bm k}^6
+\cdots+WM_{\bm k}^{2W-2}} \nonumber \\
={}& \frac{M_{\bm k}^{2W}\big( 1-M_{\bm k}^2\big)^2}{1-M_{\bm k}^{2W}
\big[1+\big(1-M_{\bm k}^2\big)W\big]}.
\end{align}
For $M_{\bm k}\ll 1$ we have simply $\zeta_0\approx M_{\bm k}^{2W}$.

\begin{figure}[tb]
\centerline{\includegraphics[width=0.9\linewidth]{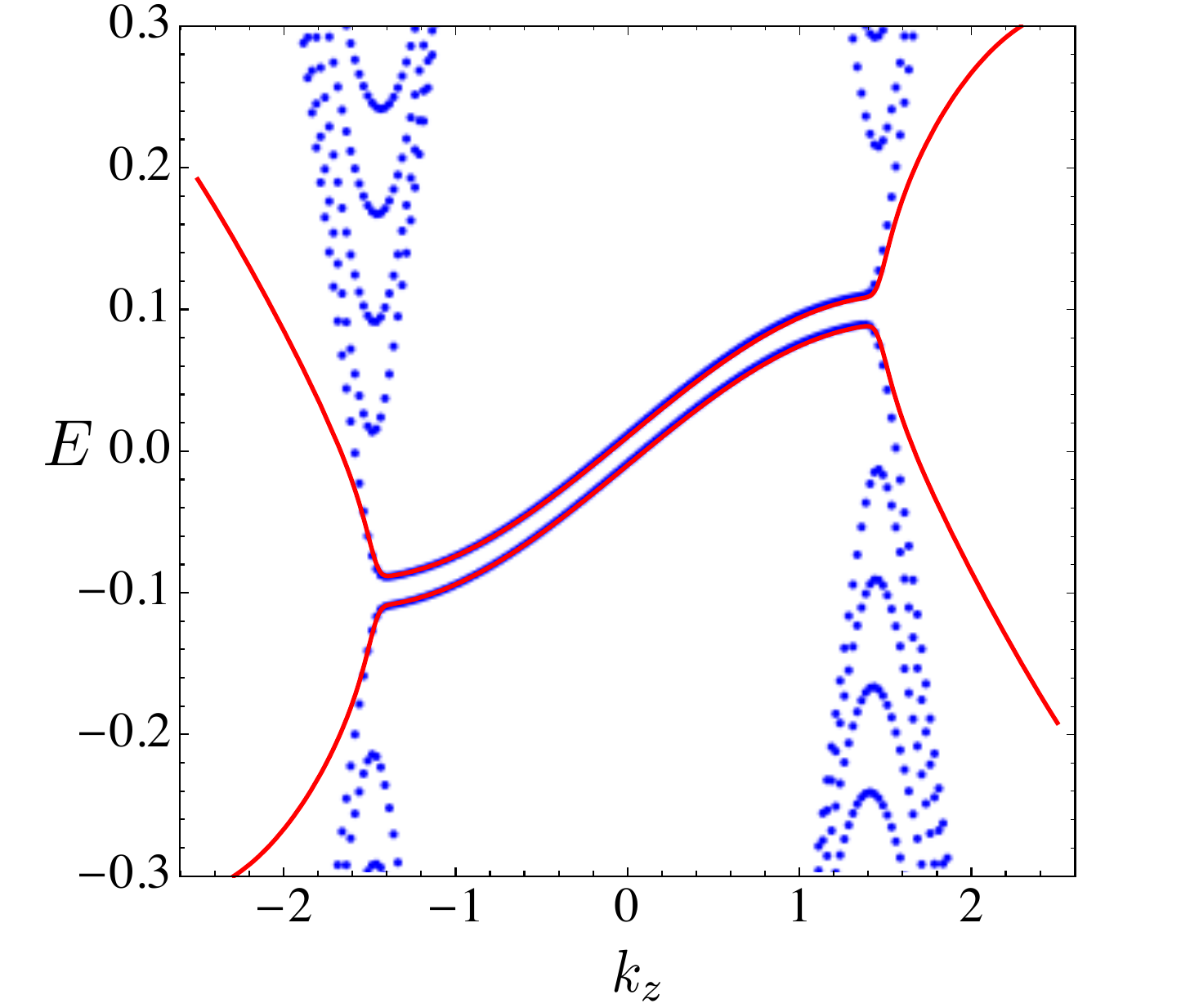}}
\caption{Dispersion relation at $k_y=0.01$ 
given by the effective Hamiltonian \eqref{Harc} (red curve),
compared to numerical results from the full Hamiltonian \eqref{Hdef} (blue dots). The parameters are the same as in Fig.\  \ref{fig_dispersion}.}
\label{fig_Heff}
\end{figure}

\begin{figure}[tb]
\centerline{\includegraphics[width=0.9\linewidth]{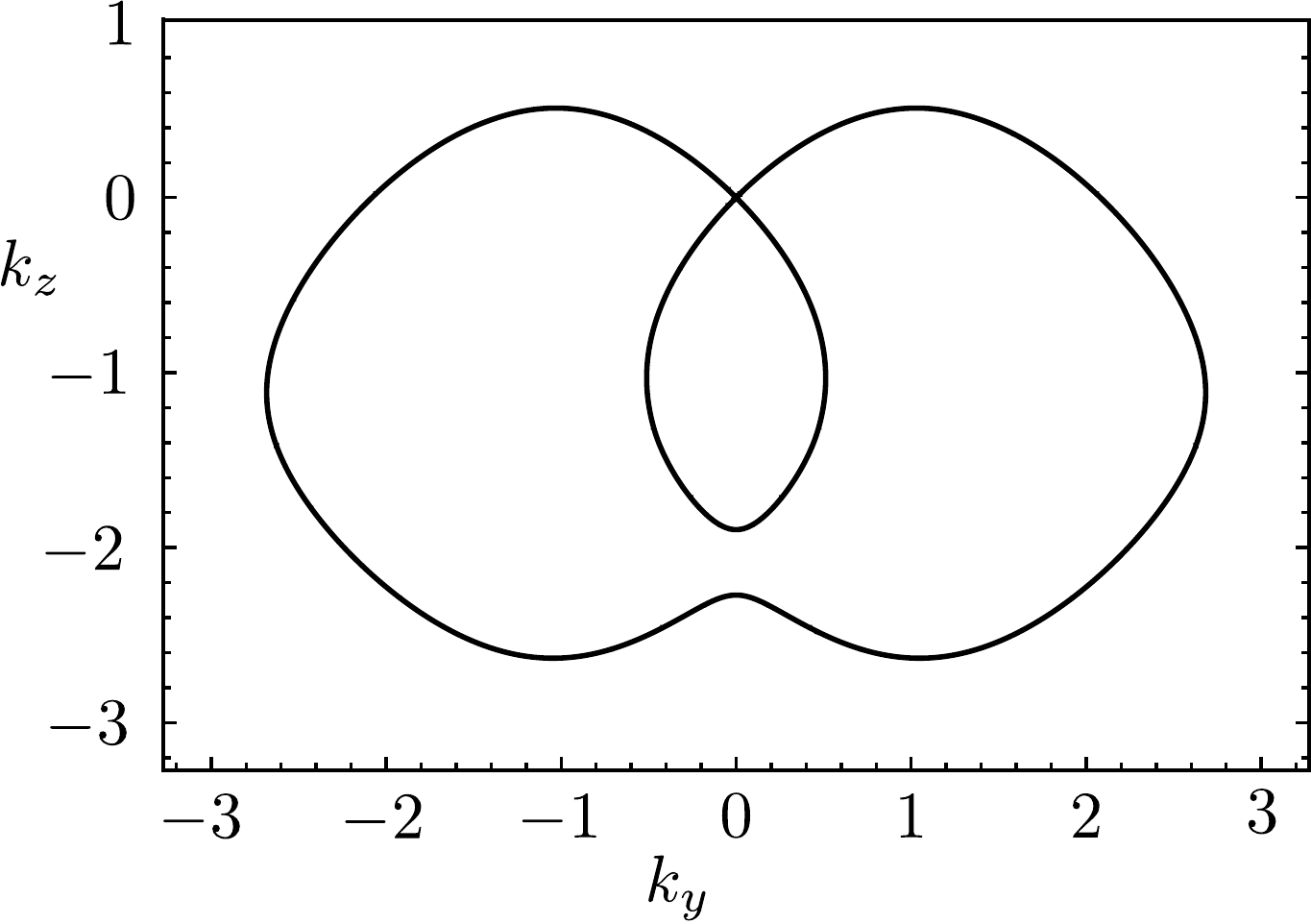}}
\caption{Fermi surface at $E=0$ with turning number $\nu=2$
given by the Hamiltonian \eqref{Hnu2}, for the parameters
$W=40$, $\beta=1.5$, $\lambda=1$, $\mu=0.6$.}
\label{fig_nu2}
\end{figure}

The corresponding effective low-energy Hamiltonian takes the form
\begin{equation}
H_\mathrm{eff} = \sigma_x \sqrt\zeta_0  + \sigma_y \sin k_y
+\lambda\sigma_0\sin k_z,
\label{Harc}
\end{equation}
where we have reinsterted the $\lambda$ term. A comparison of the energy spectrum of the effective Hamiltonian with the result from an exact numerical diagonalization of the full Hamiltonian is shown in Fig.\ \ref{fig_Heff}.

In closing, we note that a simple modification of this effective 2D Hamiltonian can be used to describe Fermi surfaces with turning number greater than unity. As an example, the Hamiltonian
\begin{equation}
\tilde{H}_\mathrm{eff} = H_\mathrm{eff}
+\mu\,(2-\cos k_z-\cos k_y)\sigma_0
\label{Hnu2}
\end{equation}
has the $\nu=2$ Fermi surface shown in Fig.\ \ref{fig_nu2}.


\begin{thebibliography}{99}
\bibitem{note3} We count occupied 2D subbands by counting the number of equi-energy contours at the Fermi energy in the $(k_y,k_z)$ plane, allowing for (nearly avoided) self-intersections. All four equi-energy contours in Fig.\ \ref{fig_Fermisurfaces} correspond to a single occupied subband.
\bibitem{note1} The generalized Lambert function ${\cal W}(t;s;a)$ is defined as the solution of the equation $e^{{\cal W}}({\cal W}-t)=a({\cal W}-s)$.
\bibitem{note2} Throughout the paper we take $\beta$ and $\lambda$ positive. The direction of motion of the edge channels indicated in Fig.\ \ref{fig_edgestates} should be inverted if either $\beta$ or $\lambda$ change sign.
\end{thebibliography}
\end{document}